# Ten steps to becoming a musculoskeletal simulation expert: A half-century of progress and outlook for the future

Scott D. Uhlrich[a], Thomas K. Uchida[b,†], Marissa R. Lee[c], and Scott L. Delp[a,c,d]

Departments of Bioengineering[a], Mechanical Engineering[c], and Orthopaedic Surgery[d]
Stanford University
Stanford, CA 94305, USA

Department of Mechanical Engineering[b]
University of Ottawa
Ottawa, ON K1N 6N5, Canada

[†]These authors have contributed equally to this work.

Please direct correspondence to:
Scott Delp, delp@stanford.edu

# Abstract

Over the past half-century, musculoskeletal simulations have deepened our knowledge of human and animal movement. This article outlines ten steps to becoming a musculoskeletal simulation expert so you can contribute to the next half-century of technical innovation and scientific discovery. We advocate looking to the past, present, and future to harness the power of simulations that seek to understand and improve mobility. Instead of presenting a comprehensive literature review, we articulate a set of ideas intended to help researchers use simulations effectively and responsibly by understanding the work on which today's musculoskeletal simulations are built, following established modeling and simulation principles, and branching out in new directions.

# Introduction

Simulations complement experimental approaches to understanding complex systems in nearly all areas of science and engineering. Simulating musculoskeletal dynamics is a powerful method for understanding the biomechanics of movement. A musculoskeletal simulation is generated by computing the motion of a musculoskeletal model that is governed by the laws of physics and the behavior of the biological system. Simulations may be driven by experimental data, a hypothesis about how an individual moves, an optimization problem, or a combination of these.

The use of simulation in biomechanics has greatly expanded over the past several decades (Fig. 1). Musculoskeletal simulations have enriched our knowledge of sport performance (Vanlandewijck et al., 2001), workplace ergonomics (Chaffin, 2001, 1969), vehicle collisions (Bose et al., 2010), and even dinosaur running (Hutchinson et al., 2005). Simulations have become so important in biomechanics that we suggest that all researchers in our field develop expertise in this area.

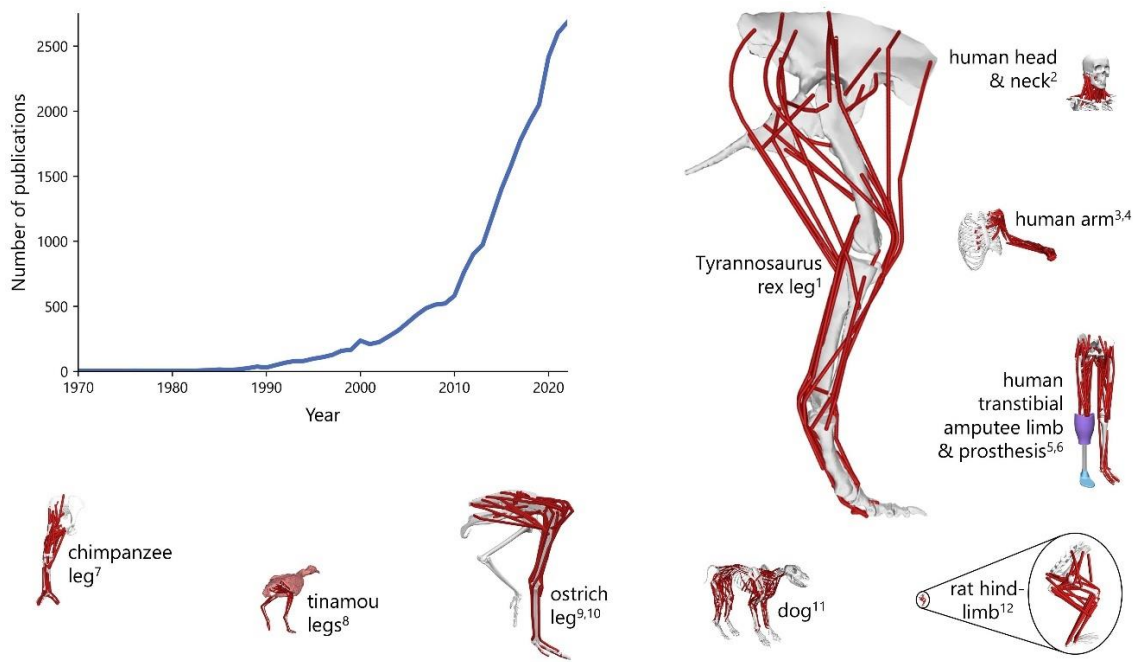

**Figure 1.** Growth of musculoskeletal modeling and simulation. Annual PubMed publications in "(musculoskeletal simulation) OR (musculoskeletal model*)" have grown by three orders of magnitude since 1970. Over the past two decades, many musculoskeletal models have been developed and shared publicly on SimTK.org for use in simulation research. Examples of shared models, shown to relative scale, have been provided by [1]Hutchinson et al. (2005), [2]Mortensen et al. (2018), [3]McFarland et al. (2019), [4]Saul et al. (2015), [5]Rajagopal et al. (2016), [6]Willson et al. (2020), [7]O'Neill et al. (2013), [8]Bishop et al. (2021b), [9]Hutchinson et al. (2015), [10]Rankin et al. (2016), [11]Stark et al. (2021), and [12]Johnson et al. (2008).

This article outlines ten steps to becoming a musculoskeletal simulation expert as we celebrate the 50th anniversary of the International Society of Biomechanics in this Special Issue. Several reviews of musculoskeletal modeling and simulation have recently been published (de Groote and Falisse, 2021; Ezati et al., 2019; Febrer-Nafría et al., 2022; Roupa et al., 2022). Thus, rather than provide another review of the literature, our goal is to outline a set of steps intended to help researchers hit the ground running with simulation research. In the first section—Look back: Know your history—we recommend three steps to understand the work upon which today's musculoskeletal simulations are built. In the second section—Look inside: Be a strong simulator—we outline three steps that every researcher should take to define and analyze their musculoskeletal simulations, even as simulation techniques advance. Finally, in the third section—Look forward: Invent the future—we offer four steps the field can take to forge new paths in musculoskeletal simulation and create lasting impact. We focus on human locomotion, but the ten steps apply to other types of human and animal movement and to musculoskeletal modeling and simulation in general.

## Look back: Know your history

## Step 1: Study early simulations

Modern musculoskeletal simulation began about fifty years ago, at roughly the same time the International Society of Biomechanics was founded. Posed with the difficulty of measuring muscle forces in vivo, researchers turned to musculoskeletal simulations to estimate muscle forces and the motions they produce. Static and dynamic optimizations of human movement were introduced in the 1970s (Chow and Jacobson, 1971; Hatze, 1976; Seireg and Arvikar, 1973). At the time, the equations of motion were derived and programmed by hand, and simulations representing just 0.5 seconds of movement could require more than twenty hours of computation by computers that filled a room. The methods used in early simulations were relatively simple compared to modern techniques, yet they paved the way for today's musculoskeletal simulations.

Several advancements have since enabled more accurate musculoskeletal simulations and deeper study of the ways in which muscles produce movement (Fig. 2). Quantitative anatomy and muscle architecture experiments in the early 1980s enabled researchers to improve model accuracy. From cadaveric limbs, Brand et al. (1982) identified three-dimensional lower-limb muscle origin and insertion coordinates. These measured coordinates were great improvements over those previously estimated from drawings in anatomy textbooks. Shortly after, Wickiewicz et al. (1983) measured lower-limb muscle properties, including muscle fiber lengths and physiological cross-sectional areas. These important studies resulted in more accurate musculoskeletal models, including the lower-extremity model developed by Hoy et al. (1990).

Through the 1980s, musculoskeletal simulations, particularly those employing dynamic optimization, were limited by barriers in computation and model development. Kane's method (Kane and Levinson, 1985) offered efficient algorithmic determination of the equations of motion and enabled the first 3D dynamic simulation of part of the gait cycle in 1990 (Yamaguchi and Zajac, 1990). Around the same time, Delp et al. (1990) developed an open-source musculoskeletal model and an interactive computer graphics interface that enabled users to manipulate musculoskeletal model parameters. This open-source model has been used in thousands of research studies, including in the groundbreaking 2001 paper describing the first 3D predictive dynamic optimization of a full cycle of walking gait (Anderson and Pandy, 2001).

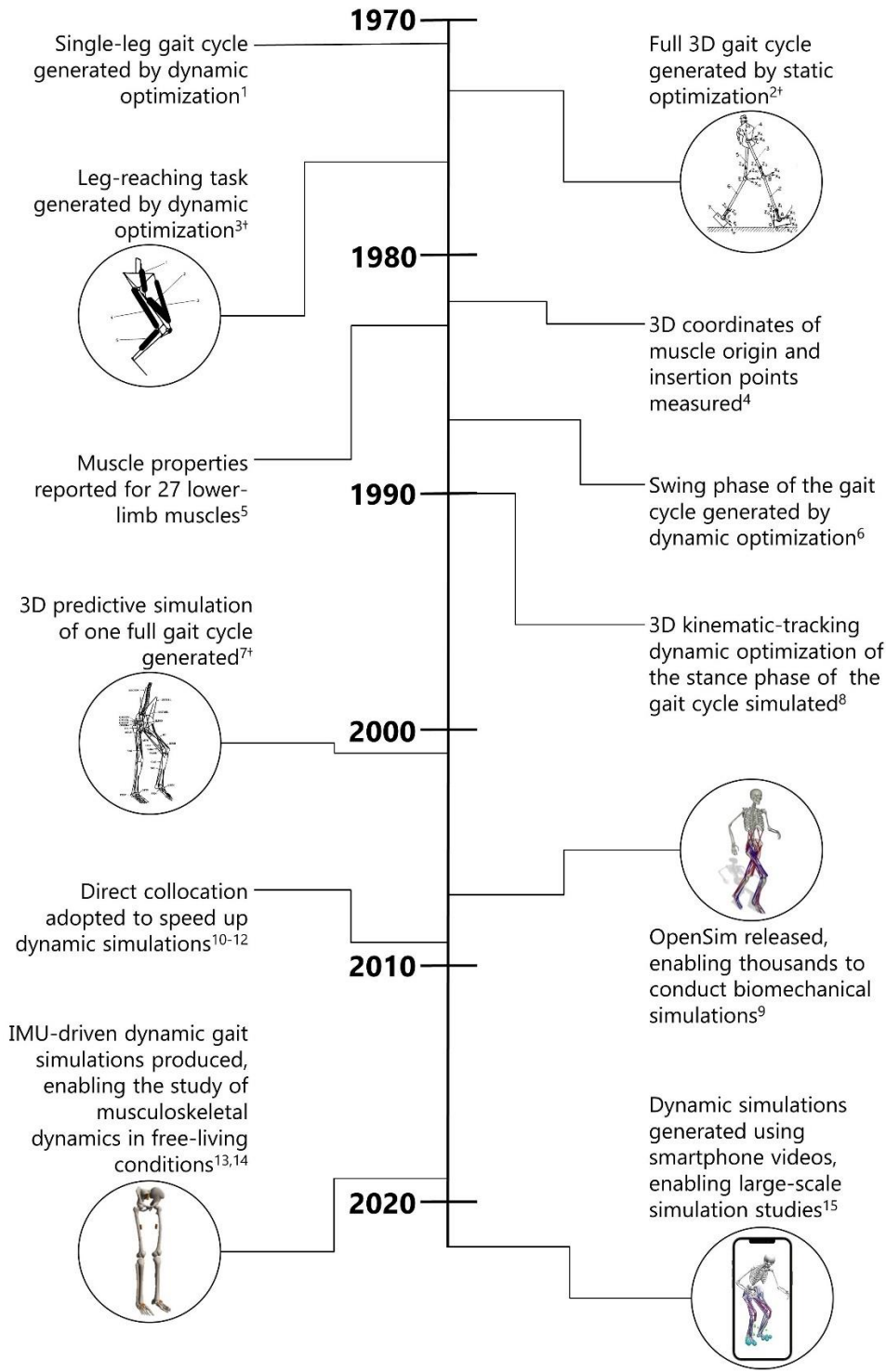

**Figure 2.** Some of the key advancements in musculoskeletal simulation over the past 50 years. [1]Chow and Jacobson (1971); [2]Seireg and Arvikar (1973); [3]Hatze (1976); [4]Brand et al. (1982); [5]Wickiewicz et al. (1983); [6]Davy and Audu (1987); [7]Yamaguchi and Zajac (1990); [8]Anderson and Pandy (2001); [9]Delp et al. (1990); [10]Ackermann and van den Bogert (2010); [11]de Groote et al. (2009); [12]Kaplan and Heegaard (2001); [13]Dorschky et al. (2019); [14]Karatsidis (2016); [15]Uhlrich, Falisse, and Kidziński et al. (2022). †Figures reprinted with permission.

In 2001, simulation of this single gait cycle required 10,000 hours of total CPU time. Efficient optimal control methods like direct collocation (Hargraves and Paris, 1987) were adopted in biomechanics in the 2000s (Ackermann and van den Bogert, 2010; de Groote et al., 2009; Kaplan and H. Heegaard, 2001). Today, researchers can perform predictive dynamic optimizations of walking in less than one hour (Dembia and Bianco et al., 2020; Falisse et al., 2019).

Advancements in quantitative anatomy, increased computational speed, and the introduction of open-source models and tools have broken barriers in musculoskeletal simulation, allowing researchers today to focus on important biomechanical questions.

## Step 2: Understand what can be learned from simulations

Muscle-driven simulations extend the insights gained from experiments. For example, it is possible to measure muscle activity, ground reaction forces, and joint motions in experiments, but it is not possible to determine how each muscle contributes to ground reaction forces and body motions with experiments alone. Muscle-driven simulations reveal the forces and motions caused by muscles and provide powerful tools for understanding muscle actions during movement. We can also leverage simulations to predict how the body responds to disease (Knarr et al., 2013; Steele et al., 2012a), surgery (Delp and Zajac, 1992), or altered muscle activations (DeMers et al., 2014). These capabilities allow us to design surgeries (Delp and Zajac, 1992) and assistive devices (Bianco et al., 2022b; Dembia et al., 2017) or simulate dangerous events, such as falls or injuries (DeMers et al., 2017).

To provide a specific example, muscle-driven simulations have revealed how muscles generate forces that support the body's weight and regulate forward progression during the stance phase of walking. We can quantify a muscle's contribution to body-weight support by determining how the muscle affects the vertical acceleration of the center of mass. Similarly, we can investigate how a muscle contributes to forward progression by analyzing its role in generating accelerations of the center of mass in the fore–aft direction (Liu et al., 2008; Neptune et al., 2001).

The same muscles that provide body-weight support during walking also regulate forward progression. For example, the vasti and gluteus maximus make important contributions to support during early stance while the vasti also reduce the body's forward velocity (Fig. 3; Liu et al., 2008). The gluteus medius provides support during midstance and, in the second half of stance, contributes to forward acceleration. The soleus and gastrocnemius contribute to vertical and forward acceleration in late stance. The contributions to body-weight support from the soleus and gastrocnemius are so important that weakness or contracture of these muscles may lead to crouch gait (Ong et al., 2019; Steele et al., 2012b). For this reason, crouch gait can be improved in some cases by wearing an ankle brace that generates a plantarflexion moment (Rosenberg and Steele, 2017).

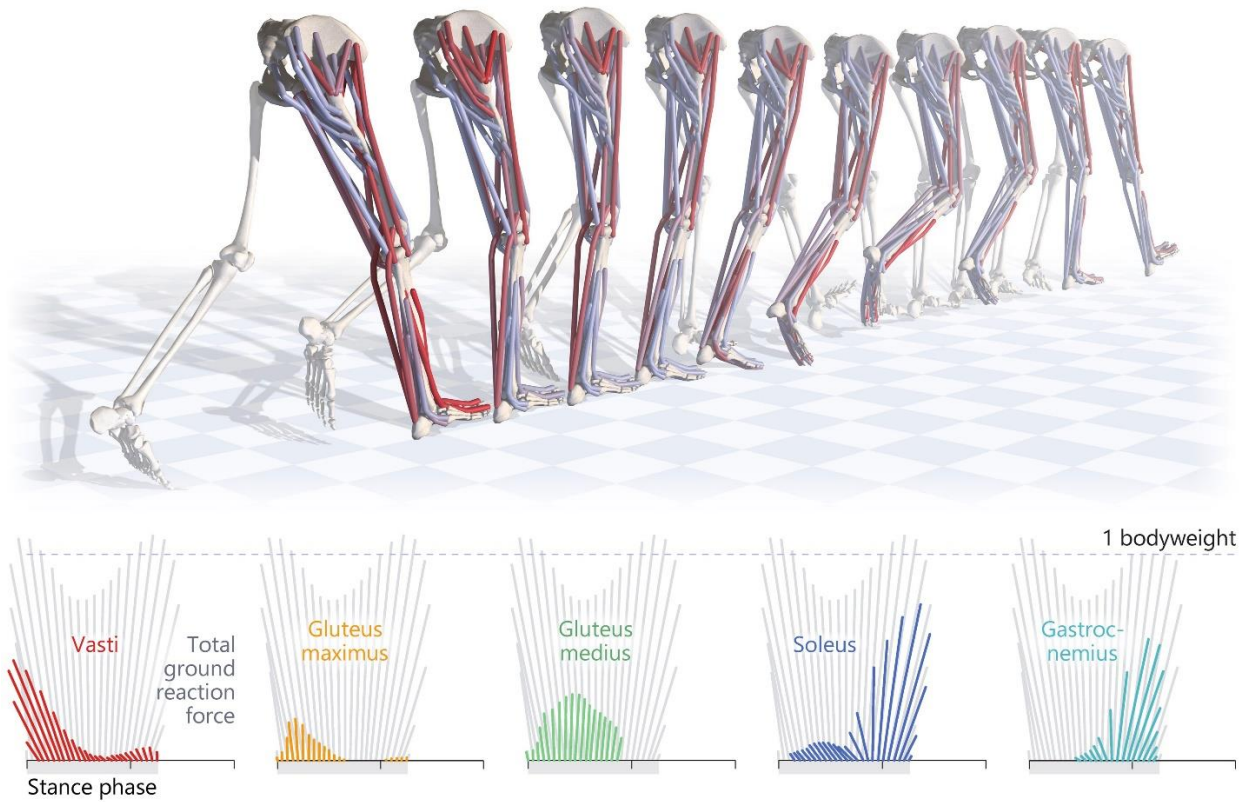

**Figure 3. Top.** Visualization of a muscle-driven simulation of walking. The colors of the muscles indicate their level of activation ranging from inactive (blue) to highly active (red). **Bottom.** Contributions of muscles to center-of-mass accelerations during the stance phase of walking. The vasti accelerate the center of mass upward and backward during early stance, while the plantarflexors accelerate the center of mass upward and forward during late stance. The gluteus maximus and gluteus medius also contribute to body-weight support. Data from Liu et al. (2008) and Dembia et al. (2017).

Step 3: Know what tools are available

Powerful musculoskeletal modeling and simulation tools are available to today's biomechanists. Some researchers produce fantastic individual contributions to the field by writing custom programs for their simulations. These programs can be fast when optimized for just one or a few applications (e.g., Mansouri and Reinbolt, 2012; van den Bogert et al., 2013). Developing custom software encourages researchers to understand every element of their code and simulations and thus requires great effort. Fortunately, since general-purpose simulation tools exist, custom software is rarely required for simulation research.

On the other end of the simulation-tool spectrum is commercial software. Commonly used commercial software in musculoskeletal modeling include the AnyBody Modeling System (AnyBody Technology, Aalborg, Denmark; Damsgaard et al., 2006), multibody dynamics simulation tools like Adams (Hexagon AB, Newport Beach, CA), and finite element tools like Abaqus (Dassault Systèmes, Vélizy-Villacoublay, France). Most motion capture companies also offer software capable of performing inverse dynamic analyses. Commercial biomechanics packages are intended to support a variety of applications. They are well tested and well suited for commercial applications, with technical support available to users. In many cases, researchers cannot access or modify the software and methodological details, which can be limiting in research.

Between custom and commercial software lies open-source software. Open-source tools in musculoskeletal simulation include OpenSim (Delp et al., 2007; Seth, Hicks, and Uchida et al., 2018) for dynamic simulations, the Calibrated EMG-Informed Neuro-musculoskeletal Modelling Toolbox (CEINMS; Pizzolato et al., 2015) for electromyography-informed simulations, Simulated Controller OptimizatioN Environment (SCONE; Geijtenbeek, 2019) for predictive simulations, and Finite Elements for Biomechanics (FEBio; Maas et al., 2012) for finite element modeling. These freely available tools invite involvement from a worldwide community of biomechanics researchers and enable reproducibility while making the underlying code available and modifiable.

It is important to know what research question you are asking so that you can determine which custom, commercial, or open-source software will provide the best answer.

# Look inside: Be a strong simulator

## Step 4: Ask the right question

A research question is the driving force behind a scientific study. To quote Carl Jung, "to ask the right question is already half the solution of a problem." Research questions that are specific and stated directly (e.g., "Does a model with weakness or contracture in the ankle plantarflexor muscles walk more slowly?" [Ong et al., 2019]) provide more direction than those that are posed more generally or are not questions at all (e.g., "The effect of muscle weakness on gait"). A good research question should also be novel, important, and interesting. Research questions in biomechanics are often posed to improve our fundamental understanding of movement (e.g., Farris and Sawicki, 2012; Umberger, 2010); to improve prevention, diagnosis, or treatment of an injury or pathology (e.g., Alentorn-Geli et al., 2009; Piazza, 2006); to enhance function and quality of life (e.g., Shull et al., 2014); to develop and disseminate analytical tools (see Section 3); and, frequently, as combinations of these (Hicks et al., 2015).

A good research question must also be answerable. Depending on the question, it may be more appropriate to perform human experiments, computer simulations, or some combination of these. One may need to collect motion data to answer the question. Alternatively, the data may already be available and can simply be aggregated from medical records (Hicks et al., 2011) or large sets of unlabeled data (Ratner et al., 2017) and analyzed to discover new insights without recruiting a single human subject. For a research question to be answerable, the study must also be feasible given the available resources, including experimental equipment, study participants, physical prototypes, musculoskeletal simulations, software, algorithms, or computational resources.

Although curiosity-driven exploration can ultimately lead to insight, scientific "dead-ends" may be avoided from the outset by imagining the goal: Supposing the desired experiments or simulations were completed, what plots would you generate and what numbers would be most important? Having a clear vision of the outcome will help to guide the study and reveal limitations in the planned approach that could be addressed by adjusting the study design. Importantly, a vision of the outcome will help to ensure that all the necessary data are collected during experiments and that any musculoskeletal models used are sufficiently detailed where the detail matters. We'll look at this in the next section.

## Step 5: Use the right tool for the job

The design of a study is driven by the research question. Inverse dynamic simulations can provide deeper understanding of experimental observations. An inverse dynamic simulation determines the values of variables in a musculoskeletal model (e.g., joint angles, net joint moments, or muscle forces over time) that best explain the observations (e.g., motion capture marker trajectories, ground reaction force data, or electromyography). Inverse dynamic simulations have been used in many human movement studies to estimate quantities that cannot be measured directly. In contrast, forward dynamic simulations use physics-based models to compute the motions that result from a set of muscle excitation patterns. For example, numerical optimization can be used to discover the muscle excitations that cause a musculoskeletal model to move to achieve a desired objective, such as minimizing metabolic cost or jumping as high as possible. Forward dynamic simulations can be useful when experimental data are unavailable—for example, when studying the locomotion of extinct animals (Bishop et al., 2021a; Hutchinson et al., 2005), potentially injurious motions (Shin et al., 2007), and assistive devices that have not yet been built (Dembia et al., 2017; Uchida et al., 2016).

All musculoskeletal simulations require a musculoskeletal model. A model is a description of reality that explains and predicts some phenomena of interest within a required precision. Any model, whether physics-based or data-driven, should be as simple as possible while capturing all the information necessary for the model to be useful (Fig. 4). Increasing model complexity can introduce new parameters and uncertainties, increase computation time, and obscure the information required to answer the research question. On the other hand, overly simple models may embed assumptions that lead to inaccurate conclusions. Whether computational or "purely" experimental, nearly every biomechanics study uses a model of some kind. For example, calculation of the knee flexion angle during gait may use a model that makes assumptions about the possible motions of the knee or the relationships of motion capture markers to the underlying bones. Some typical questions to consider when choosing or developing a model to answer a research question include which degrees of freedom the model should have, whether muscles are required, and how contact will be represented (e.g., using measurements, kinematic constraints, a contact model, or machine learning). Many musculoskeletal models have been built, validated, and openly shared for the biomechanics community to use (Fig. 1).

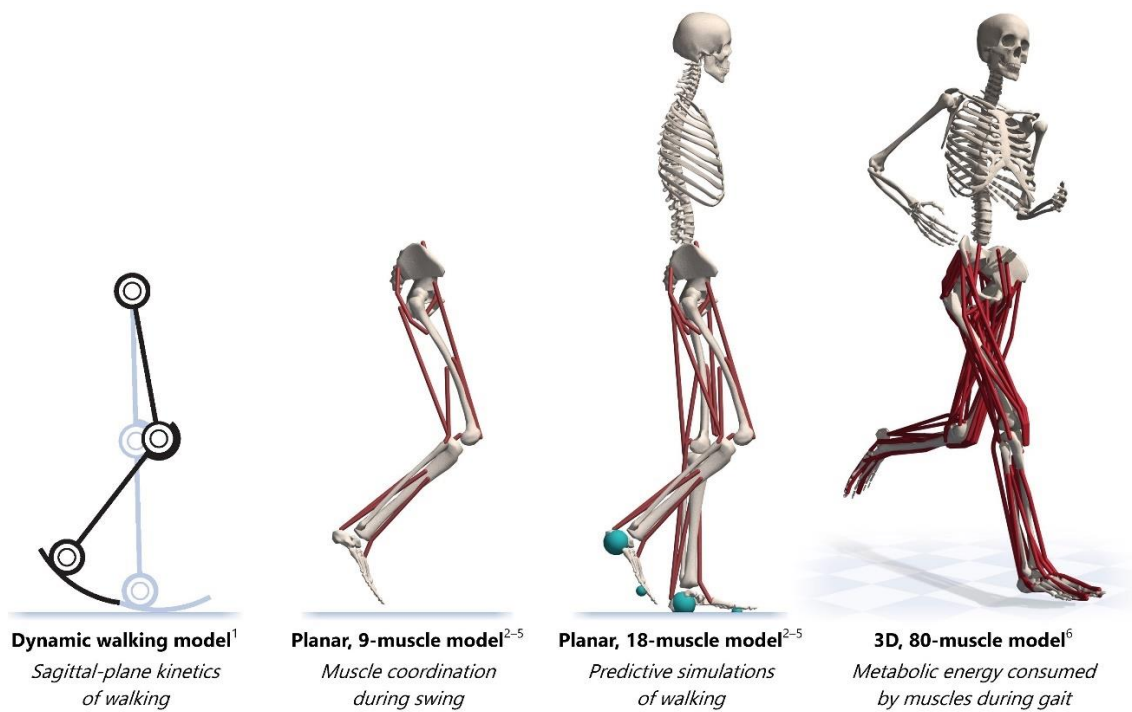

**Figure 4.** Four gait models of increasing complexity and an example of a study for which each model might be appropriate. The research question dictates the requirements for a musculoskeletal model. [1]Kuo and Donelan (2010); [2]Delp et al. (1990); [3]Yamaguchi and Zajac (1989); [4]Anderson and Pandy (1999); [5]Anderson and Pandy (2001); [6]Rajagopal et al. (2016).

## Step 6: Preserve a healthy skepticism

To quote Carl Sagan, "extraordinary claims require extraordinary evidence." This so-called "Sagan standard" reminds us to aspire to be skeptical, unbiased scientists. From simulations, we often seek general conclusions about human movement. The robustness of these conclusions must be tested by exploring simulation results over a range of operating conditions, performing sensitivity analyses, and evaluating conclusions with real-world data. Uncertainties in the variables of interest should be evaluated and reported. The most reliable conclusions are those that have been found consistently using many datasets, models, simulation strategies, and software packages.

Models and simulations must be verified and validated. Verification tests ensure that simulation results satisfy known relationships, such as energy conservation and Newton's laws. Validation tests ensure that simulation results agree with real-world observations. For example, even if a simulation passed verification tests, a result indicating that the cost of transport decreases as walking speed increases from self-selected speed would contradict established principles of human locomotion. Hicks et al. (2015) provide an overview of verification and validation for modeling and simulation studies as well as recommendations.

It is important to be aware of the limitations of models and simulation tools that you employ. Modelers should identify and quantify the largest sources of error in their analyses and evaluate how these errors affect the outputs of interest. For example, OpenSim uses residual actuators to ensure that Newton's second law is satisfied despite dynamic inconsistencies between measured forces $F$, estimated segment masses $m$, and measured accelerations $a$. The Residual Reduction Algorithm in OpenSim aims to minimize the "residual forces" $F_{\text{residual}}$:

$$F + F_{\text{residual}} = ma$$

Some, but not all, software tools quantify and report the errors in the underlying models. Whether reported or not, modeling errors always exist and must be considered when drawing conclusions. It is also important to know the limitations of the musculoskeletal model you are using to ensure the model is operating within the range for which it has been validated. Additional validation may be required to ensure the model is suitable for your study. For example, the gait model developed by Rajagopal et al. (2016) was extended by Lai et al. (2017) to model movements involving substantial hip and knee flexion. Errors also appear during simulations—for example, integration tolerances and interpolation between data points pollute calculations with numerical error. Convergence analyses can be used to determine simulation parameters such as integration tolerances and termination criteria for iterative algorithms that maintain sufficiently small errors.

# Look forward: Invent the future

## Step 7: Embrace new techniques

Leveraging technical advances from other fields can catalyze progress in musculoskeletal simulation. For example, direct collocation, which originated in aerospace engineering (Hargraves and Paris, 1987), has made it far easier to generate predictive simulations using complex musculoskeletal models (Falisse et al., 2019). Another emerging trend is applying machine learning to biomechanics problems (Halilaj et al., 2018). Understanding when to apply physics-based models versus machine learning models and devising clever ways to combine the two are becoming important skills in computational biomechanics.

Physics-based simulations are usually more generally applicable and interpretable than machine learning models. If we have a good model, we can estimate unmeasurable quantities based on established physical laws. Improvements to the fidelity of these models (e.g., through the Knee Grand Challenge [Fregly et al., 2012; Kinney et al., 2013]) can have widespread benefits for other modeling problems. Two drawbacks of the physics-based approach are that the questions we can answer are limited by how well we model the physical system, and our results can be sensitive to assumptions, some of which may be difficult to test.

Machine learning models are well suited to approximate phenomena for which we lack good physical models, enabling accurate predictions when given input data that are within the distribution of the training dataset. For example, machine learning has been used to predict fall risk (Tunca et al., 2020), recognize intent for prosthesis control (Labarrière et al., 2020), and detect freezing of gait in Parkinson's patients (O'Day, Lee, and Seagers et al., 2022). Common drawbacks of this approach are that models typically require large amounts of training data, are often not interpretable, struggle to generalize beyond the distribution of the training dataset, and do not take advantage of our knowledge of the physical system.

Merging physics-based and machine learning approaches can yield more accurate models that are trained on smaller datasets (Fig. 5). One common approach is to use musculoskeletal models to generate training data for machine learning models that use sparse inputs, such as data from a small number of inertial measurement units or acoustic emissions from a joint (Jiang et al., 2022; Scherpereel et al., 2021; Sharifi Renani et al., 2021; Uhlrich, Falisse, and Kidziński et al., 2022). Another approach is to build elements of the physical system into the machine learning model. For example, physics simulations can improve the accuracy of a deep-learning-based pose estimation model (Shimada et al., 2020; Yuan et al., 2021), and neural networks can be regularized with physics-based terms in the loss function (Raissi et al., 2019; Zhang et al., 2023). Yet another approach is to use physical models for elements of the system that we understand well and to train machine learning models for the remaining elements. Using reinforcement learning, for example, a network can learn a model of sensory-motor control (which is challenging to model mechanistically) that enables a musculoskeletal model to navigate complex environments (Kidziński et al., 2018; Song et al., 2021). Novel methods for combining physics and data into unified models will likely be a popular theme in future biomechanics research.

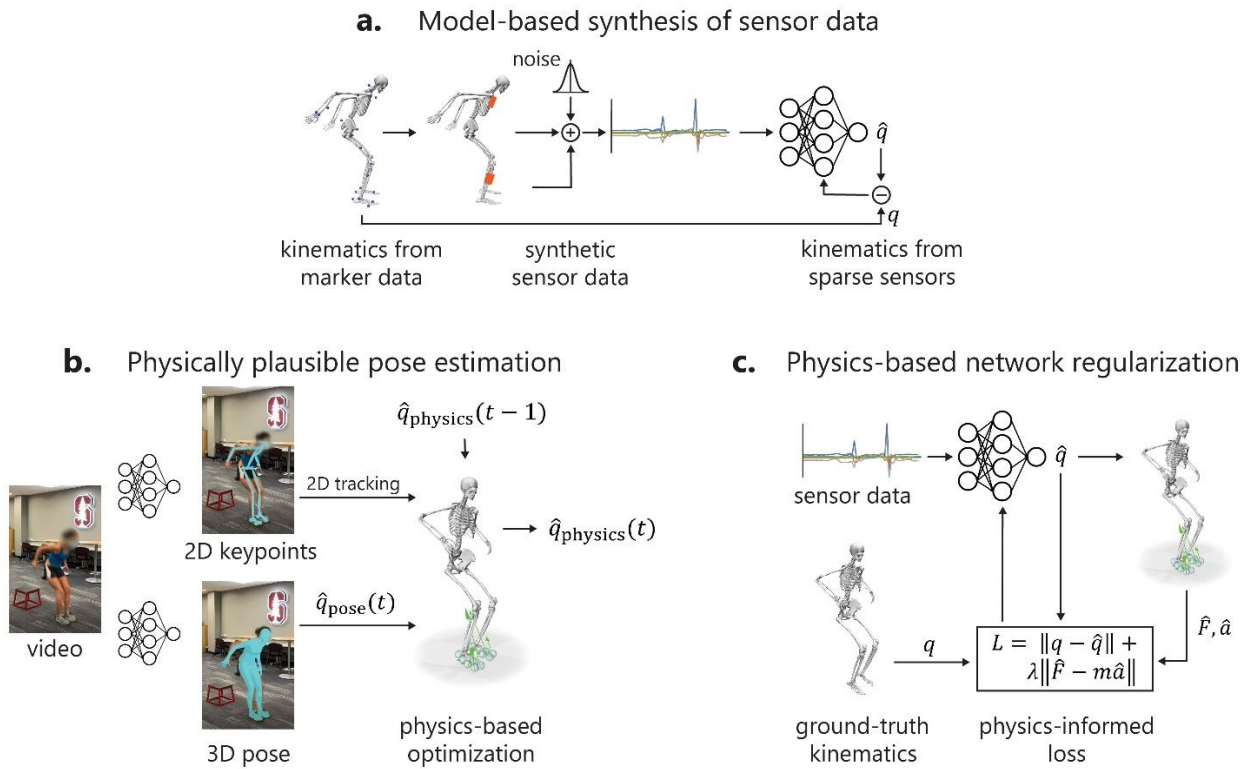

**Figure 5.** Approaches to combining physics-based modeling and machine learning. **(a)** Musculoskeletal models can be used to synthesize inputs to machine learning models (Jiang et al., 2022; Sharifi Renani et al., 2021; Uhlrich, Falisse, and Kidziński et al., 2022). Here, $q$ and $\hat{q}$ represent the ground-truth and predicted generalized coordinates of a musculoskeletal model. **(b)** Physics-based optimization can improve the accuracy of deep-learning-based pose estimation (Shimada et al., 2020; Yuan et al., 2021). **(c)** Physics terms can regularize the loss function of a deep learning model, helping to reduce overfitting (Zhang et al., 2023). In this simple example, $F$ represents ground reaction forces, $a$ represents center-of-mass accelerations, and $\lambda$ changes the relative weighting between the tracking error and physics terms in the loss function. Fully connected neural networks are used to depict machine learning models of any architecture.

## Step 8: Leverage simulation at scale

Improvements in computational efficiency and the release of easy-to-use software packages continue to expand the community that has access to musculoskeletal simulation (Fig. 1). Despite these advances, most inverse dynamic simulation studies continue to include only a small number of participants, in part due to the time required to collect and process optical motion capture data. Use of optical motion capture and force plates also limits most studies to the controlled laboratory setting. Mobile sensors reduce these barriers to simulation studies that include hundreds of participants. Simulations can be generated from data collected outside the lab using sensors that are orders of magnitude less expensive than marker-based motion capture, such as standard video cameras or inertial measurement units (al Borno and O'Day et al., 2022; Dorschky et al., 2019; Haralabidis et al., 2020; Karatsidis et al., 2016; Slade et al., 2022; Uhlrich, Falisse, and Kidziński et al., 2022). We anticipate that automated pipelines for generating simulations from mobile sensor data will enable large studies with sufficient statistical power to discover movement biomarkers that can be easily measured in the clinic or home.

We envision a future where a digital representation of our musculoskeletal system is continuously monitored, allowing for the prediction and prevention of musculoskeletal injury and disease. Our digital twins (Glaessgen and Stargel, 2012; Pizzolato et al., 2019) will be created passively from videos of us moving through our natural environments or from wearable sensors embedded into our clothing. With these technologies, we will capture the mechanics of events, such as falls and sports injuries, that cannot be captured in a laboratory. With sufficient data, we will then be able to predict these events and deliver just-in-time interventions. For example, sensor-embedded clothing continuously estimating the dynamics of an older adult could predict a future loss of balance and command a balance-restoring exoskeleton torque (Bianco et al., 2022a). Furthermore, inexpensively generated simulations will lead to a better understanding of the role of musculoskeletal mechanics in the onset of movement-related diseases. Large-scale observational studies, such as the UK Biobank or the Osteoarthritis Initiative, will be able to incorporate estimates of musculoskeletal dynamics with the same ease as step counts, which are commonly measured in population health studies. Instead of simulating seconds of movement, future simulations will leverage these rich longitudinal datasets and simulate years into the future, driving the creation of personalized preventative interventions.

## Step 9: Tackle the hard problems

Many challenges remain in modeling the neuro-musculoskeletal system. First, we need to integrate better models of neural control into musculoskeletal models. Current optimization-based approaches for estimating muscle excitations can generate realistic motion, but they are insufficient for studying the impact of pain or neurological pathology on motor control and movement. Next, we need better models of temporal changes in neural control and the musculoskeletal system. Improved models of how motor control changes over time will elucidate how humans acquire new skills and adapt to assistive devices. Enhanced models of the interaction between long-duration tissue loading and biological responses will help us study soft tissue injuries, tissue remodeling, and muscle fatigue. More generally, solutions to complex, movement-related health challenges, such as the global inactivity pandemic, will require multidisciplinary collaborations that incorporate biomechanics, psychology, sociology, and environmental factors (Althoff et al., 2017; Crum and Langer, 2007; Hicks et al., 2023; King et al., 2020).

Solving these and other grand challenges will require increasingly multidisciplinary training and teams. Bridging biomechanics, neuroscience, computer science, robotics, and psychology will lead to breakthrough technologies that improve mobility. For example, using musculoskeletal models to inform prosthesis control (Sartori et al., 2018), neural stimulation (Angeli et al., 2018), or rehabilitation robot control (Pizzolato et al., 2019) could lead to dramatic functional improvements for individuals with amputation or neurological injury. Additionally, blurring the boundaries between simulations and experiments will improve our models and help translate simulation-based insights. The design of joint-offloading gait modifications exemplifies this process. Fregly et al. (2007) used a torque-driven simulation to design a gait modification strategy that reduced the knee adduction moment, a surrogate measure of medial knee loading. Walter et al. (2010) tested this gait modification experimentally and found that increases in muscle force (which were not modeled by Fregly et al. [2007]) attenuated the expected reductions in medial knee contact force. These findings inspired the development of simulations to design interventions that reduce the muscle contribution to knee loading (DeMers et al., 2014; van Veen et al., 2019), which were later tested experimentally (Uhlrich et al., 2022). This positive feedback between simulations and experiments will accelerate as real-time simulations become common (Pizzolato et al., 2017; Stanev et al., 2021; van den Bogert et al., 2013).

The ability to solve problems with global impact will be enhanced by creating global teams with diverse members. Since simulation research can be conducted without a laboratory, it is conceivable that the modeling and simulation community could be more globally diverse than experimental domains that require specialized equipment. If this were true, we might expect the OpenSim community to be more representative of the global population than the International Society of Biomechanics (ISB) membership, which includes both computational and experimental researchers. However, both the ISB and OpenSim communities are concentrated in Europe and North America, and researchers from Asia, Africa, and South America are underrepresented in both communities (Fig. 6). While these disparities are multifactorial, open-source code and datasets will continue to lower the barrier to entry for simulation research. In addition, assembling international teams to solve problems of mutual interest will enable bidirectional sharing of expertise (Haelewaters et al., 2021). To this end, we have found "virtual research office hours," during which we consult with research teams from around the world, to be an effective and inclusive way to disseminate simulation knowledge and to learn from research teams based far from our own.

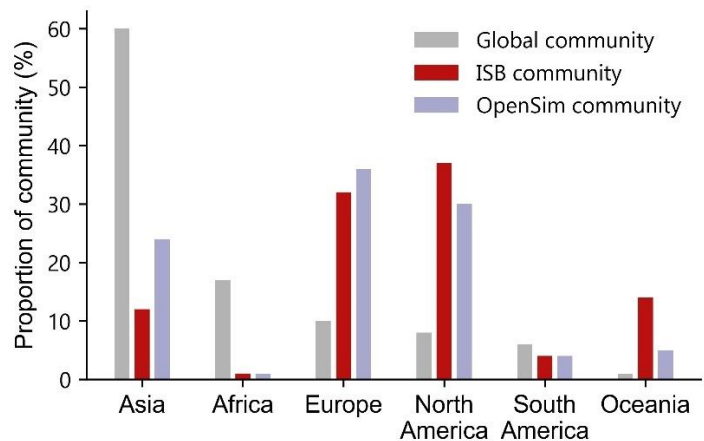

**Figure 6.** The distribution across continents of the International Society of Biomechanics (ISB) membership and the OpenSim community compared to the global population. Relative to their proportion of the global population, members of ISB and the OpenSim community from Europe, North America, and Oceania are over-represented, while those from Asia, South America, and Africa are under-represented. Data from https://isbweb.org and visits to OpenSim documentation (https://simtk-confluence.stanford.edu:8443/display/OpenSim) as of December 2022.

Step 10: Make an impactful contribution

Simulations have led to clinically relevant research insights in the past two decades, and they will likely have more direct impact on clinical care in the near future (Fig. 7; Killen et al., 2020). Simulations have elucidated general principles that can inform care, such as the relative tissue loads induced by common rehabilitation exercises (Pellikaan et al., 2018; van Rossom et al., 2018). Although simulations rarely inform patient-specific treatment decisions, easy-to-use tools to inform surgical planning with simulation-based insights (Pitto et al., 2019; Rajagopal et al., 2020) and tools to quickly generate simulations using portable sensors demonstrate progress toward this long-standing goal (Stanev et al., 2021; Uhlrich, Falisse, and Kidziński et al., 2022). As technical advances facilitate the integration of simulations into clinical workflows, it is critical for researchers to spend time with clinicians and patients to understand unmet needs: Which decisions are data-poor? Which generic treatments have poor outcomes? What are the stakeholder incentives and constraints? As simulations begin to inform device design, intervention design, and treatment decisions, it will be important to demonstrate efficacy through clinical trials. For example, van Rossom et al. (2018) used simulated knee contact forces to propose an ordered introduction of rehabilitation exercises following anterior cruciate ligament reconstruction. A randomized controlled trial demonstrating that this simulation-informed rehabilitation protocol slows the cartilage degeneration that is common after ligament injury (Monu et al., 2017) could lead to rapid clinical adoption.

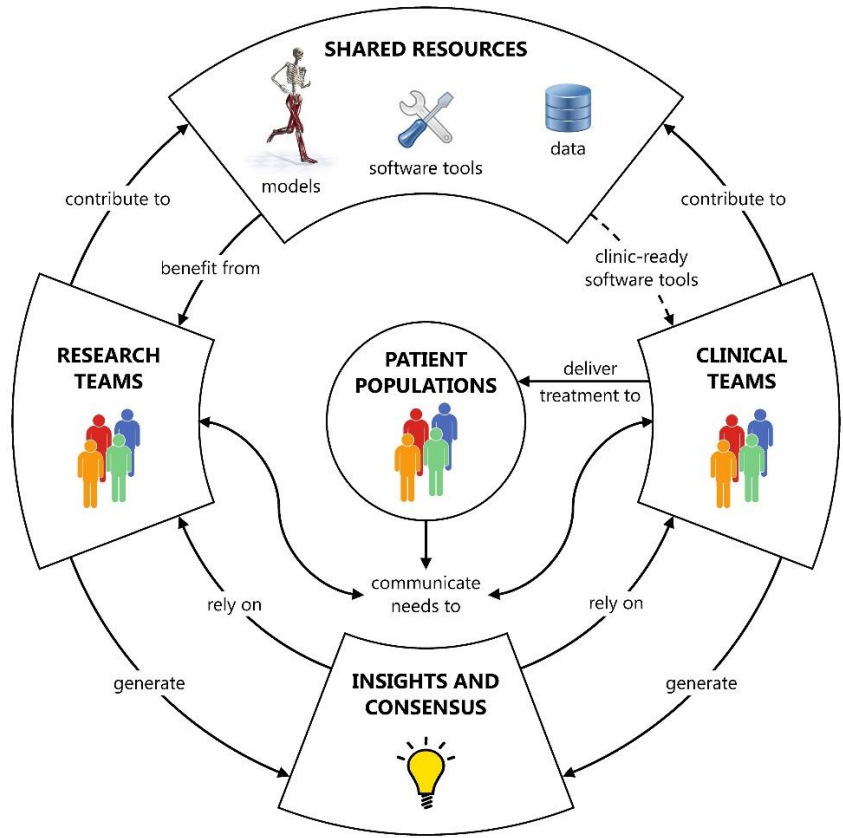

**Figure 7.** Clinically relevant research questions are driven by the needs of patient populations. Research teams and clinical teams work together to develop shared resources and to arrive at generalizable insights and consensus. In the future, use of software tools to improve clinical treatment (dashed line) will be ubiquitous.

Sharing code and models along with publications increases impact. Although it is time consuming, preparing code for sharing adds a useful verification step, improves reproducibility, and enhances confidence in results. Others may find errors in your shared code—this advances the field. Making models and simulation software available allows others to build on your work. When possible, building custom algorithms that extend widely used platforms will increase impact (e.g., CEINMS [Pizzolato et al., 2015], SCONE [Geijtenbeek, 2019], or OpenSim Moco [Dembia and Bianco et al., 2020]), but doing so requires maintenance. Increasing funding and career incentives for releasing, documenting, and maintaining open-source software will increase the number of valuable simulation tools available to the field.

Sharing data is another way to have impact. To train generalizable machine learning models, we need publicly available biomechanics datasets that are orders of magnitude larger than what currently exists. ImageNet, a dataset of images that catalyzed profound developments in deep learning, contains 3.2 million samples (Deng et al., 2009). In contrast, AMASS, the largest publicly available motion database, contains only 11,000 motions (Mahmood et al., 2019). The biomechanics community can create large datasets if we (1) make a practice of including data-sharing options in consent documents for experiments involving human subjects, (2) understand the benefits to the field and individual scientists (e.g., higher citation rates [Colavizza et al., 2020]), and (3) leverage tools that simplify data collection and sharing (e.g., addBiomechanics [Werling et al., 2022]). Representing motion data in terms of common musculoskeletal models will help standardize data across sensing modalities.

## Conclusions

Early pioneers of musculoskeletal simulation may have found it difficult to believe that, in 50 years, one would be able to download a musculoskeletal model created by a colleague across the world and run a dynamic simulation in less than an hour without writing a single equation. Today's biomechanists have access to dozens of validated models and software tools to generate forward and inverse dynamic simulations. The volume of movement data available to biomechanists is increasing rapidly. Identifying clever ways to integrate our biomechanical knowledge of the neuro-musculoskeletal system with data-driven approaches will be key to developing models that predict and improve health.

Biomechanists of the 1970s also likely did not predict that a telephone would be able to generate input data for a simulation. Technical advances in the next 50 years are sure to surprise us, but even today's technology is poised to have great impact. By the 100th anniversary of the International Society of Biomechanics, we predict that insights from simulations will have contributed to solving today's pressing healthcare challenges, such as reducing the incidence of falls; improving mobility following a stroke, an amputation, or a spinal cord injury; reducing rates of injuries; and reducing the prevalence of osteoarthritis.

We live in an exciting time of musculoskeletal simulation research. By leveraging insights from the decades of previous work in the field, following established research principles, and adopting open, forward-thinking mindsets, discoveries in biomechanics will continue to improve our lives.

## Acknowledgements

This work was supported by the National Institutes of Health (grants P41EB027060, P2CCHD101913, and R01GM124443) the Wu Tsai Human Performance Alliance, and Stanford Data Science. The study sponsors had no involvement in the study design; in the collection, analysis, or interpretation of data; in the writing of the manuscript; or in the decision to submit the manuscript for publication.

# References

Ackermann, M., van den Bogert, A.J., 2010. Optimality principles for model-based prediction of human gait. Journal of Biomechanics 43, 1055–1060. https://doi.org/10.1016/j.jbiomech.2009.12.012

al Borno, M., O'Day, J., Ibarra, V., Dunne, J., Seth, A., Habib, A., Ong, C., Hicks, J., Uhlrich, S., Delp, S., 2022. OpenSense: An open-source toolbox for inertial-measurement-unit-based measurement of lower extremity kinematics over long durations. Journal of NeuroEngineering and Rehabilitation 19, 22. https://doi.org/10.1186/s12984-022-01001-x

Alentorn-Geli, E., Myer, G.D., Silvers, H.J., Samitier, G., Romero, D., Lázaro-Haro, C., Cugat, R., 2009. Prevention of non-contact anterior cruciate ligament injuries in soccer players. Part 1: Mechanisms of injury and underlying risk factors. Knee Surgery, Sports Traumatology, Arthroscopy 17, 705–729. https://doi.org/10.1007/s00167-009-0813-1

Althoff, T., Sosič, R., Hicks, J.L., King, A.C., Delp, S.L., Leskovec, J., 2017. Large-scale physical activity data reveal worldwide activity inequality. Nature 547, 336–339. https://doi.org/10.1038/nature23018

Anderson, F.C., Pandy, M.G., 1999. A dynamic optimization solution for vertical jumping in three dimensions. Computer Methods in Biomechanics and Biomedical Engineering 2, 201–231. https://doi.org/10.1080/10255849908907988

Anderson, F.C., Pandy, M.G., 2001. Dynamic optimization of human walking. Journal of Biomechanical Engineering 123, 381–390. https://doi.org/10.1115/1.1392310

Angeli, C.A., Boakye, M., Morton, R.A., Vogt, J., Benton, K., Chen, Y., Ferreira, C.K., Harkema, S.J., 2018. Recovery of over-ground walking after chronic motor complete spinal cord injury. New England Journal of Medicine 379, 1244–1250. https://doi.org/10.1056/NEJMoa1803588

Bianco, N.A., Collins, S.H., Liu, K., Delp, S.L., 2022a. Simulating the effect of ankle plantarflexion and inversion-eversion exoskeleton torques on center of mass kinematics during walking. bioRxiv. https://doi.org/10.1101/2022.11.07.515398

Bianco, N.A., Franks, P.W., Hicks, J.L., Delp, S.L., 2022b. Coupled exoskeleton assistance simplifies control and maintains metabolic benefits: A simulation study. PLOS ONE 17, e0261318. https://doi.org/10.1371/journal.pone.0261318

Bishop, P.J., Cuff, A.R., Hutchinson, J.R., 2021a. How to build a dinosaur: Musculoskeletal modeling and simulation of locomotor biomechanics in extinct animals. Paleobiology 47, 1–38. https://doi.org/10.1017/pab.2020.46

Bishop, P.J., Michel, K.B., Falisse, A., Cuff, A.R., Allen, V.R., de Groote, F., Hutchinson, J.R., 2021b. Computational modelling of muscle fibre operating ranges in the hindlimb of a small ground bird (*Eudromia elegans*), with implications for modelling locomotion in extinct species. PLOS Computational Biology 17, e1008843. https://doi.org/10.1371/journal.pcbi.1008843

Bose, D., Crandall, J.R., Untaroiu, C.D., Maslen, E.H., 2010. Influence of pre-collision occupant parameters on injury outcome in a frontal collision. Accident Analysis & Prevention 42, 1398–1407. https://doi.org/10.1016/j.aap.2010.03.004

Brand, R.A., Crowninshield, R.D., Wittstock, C.E., Pedersen, D.R., Clark, C.R., van Krieken, F.M., 1982. A model of lower extremity muscular anatomy. Journal of Biomechanical Engineering 104, 304–310. https://doi.org/10.1115/1.3138363

Chaffin, D., 2001. Digital Human Modeling for Vehicle and Workplace Design. Society of Automotive Engineers, Inc., Warrendale, PA.

Chaffin, D.B., 1969. A computerized biomechanical model—Development of and use in studying gross body actions. Journal of Biomechanics 2, 429–441. https://doi.org/10.1016/0021-9290(69)90018-9

Chow, C.K., Jacobson, D.H., 1971. Studies of human locomotion via optimal programming. Mathematical Biosciences 10, 239–306. https://doi.org/10.1016/0025-5564(71)90062-9

Colavizza, G., Hrynaszkiewicz, I., Staden, I., Whitaker, K., McGillivray, B., 2020. The citation advantage of linking publications to research data. PLOS ONE 15, e0230416. https://doi.org/10.1371/journal.pone.0230416

Crum, A.J., Langer, E.J., 2007. Mind-set matters. Psychological Sciences 18, 165–171. https://doi.org/10.1111/j.1467-9280.2007.01867.x

Damsgaard, M., Rasmussen, J., Christensen, S.T., Surma, E., de Zee, M., 2006. Analysis of musculoskeletal systems in the AnyBody Modeling System. Simulation Modelling Practice and Theory 14, 1100–1111. https://doi.org/10.1016/j.simpat.2006.09.001

Davy, D.T., Audu, M.L., 1987. A dynamic optimization technique for predicting muscle forces in the swing phase of gait. Journal of Biomechanics 20, 187–201. https://doi.org/10.1016/0021-9290(87)90310-1

de Groote, F., Falisse, A., 2021. Perspective on musculoskeletal modelling and predictive simulations of human movement to assess the neuromechanics of gait. Proceedings of the Royal Society B: Biological Sciences 288, 20202432. https://doi.org/10.1098/rspb.2020.2432

de Groote, F., Pipeleers, G., Jonkers, I., Demeulenaere, B., Patten, C., Swevers, J., de Schutter, J., 2009. A physiology based inverse dynamic analysis of human gait: Potential and perspectives. Computational Methods in Biomechanics and Biomedical Engineering 12, 563–574. https://doi.org/10.1080/10255840902788587

Delp, S.L., Anderson, F.C., Arnold, A.S., Loan, P., Habib, A., John, C.T., et al., 2007. OpenSim: Open-source software to create and analyze dynamic simulations of movement. IEEE Transactions on Biomedical Engineering 54, 1940–1950. https://doi.org/10.1109/TBME.2007.901024

Delp, S.L., Loan, J.P., Hoy, M.G., Zajac, F.E., Topp, E.L., Rosen, J.M., 1990. An interactive graphics-based model of the lower extremity to study orthopaedic surgical procedures. IEEE Transactions on Biomedical Engineering 37, 757–767. https://doi.org/10.1109/10.102791

Delp, S.L., Zajac, F.E., 1992. Force- and moment-generating capacity of lower-extremity muscles before and after tendon lengthening. Clinical Orthopaedics and Related Research 284, 247–259. https://doi.org/10.1097/00003086-199211000-00035

Dembia, C.L., Bianco, N.A., Falisse, A., Hicks, J.L., Delp, S.L., 2020. OpenSim Moco: Musculoskeletal optimal control. PLOS Computational Biology 16, e1008493. https://doi.org/10.1371/journal.pcbi.1008493

Dembia, C.L., Silder, A., Uchida, T.K., Hicks, J.L., Delp, S.L., 2017. Simulating ideal assistive devices to reduce the metabolic cost of walking with heavy loads. PLOS ONE 12, e0180320. https://doi.org/10.1371/journal.pone.0180320

DeMers, M.S., Hicks, J.L., Delp, S.L., 2017. Preparatory co-activation of the ankle muscles may prevent ankle inversion injuries. Journal of Biomechanics 52, 17–23. https://doi.org/10.1016/j.jbiomech.2016.11.002

DeMers, M.S., Pal, S., Delp, S.L., 2014. Changes in tibiofemoral forces due to variations in muscle activity during walking. Journal of Orthopaedic Research 32, 769–776. https://doi.org/10.1002/jor.22601

Deng, J., Dong, W., Socher, R., Li, L.-J., Li, K., Fei-Fei, L., 2009. ImageNet: A large-scale hierarchical image database. In Proceedings of the 2009 IEEE Conference on Computer Vision and Pattern Recognition. Miami. https://doi.org/10.1109/CVPR.2009.5206848

Dorschky, E., Nitschke, M., Seifer, A.-K., van den Bogert, A.J., Eskofier, B.M., 2019. Estimation of gait kinematics and kinetics from inertial sensor data using optimal control of musculoskeletal models. Journal of Biomechanics 95, 109278. https://doi.org/10.1016/j.jbiomech.2019.07.022

Ezati, M., Ghannadi, B., McPhee, J., 2019. A review of simulation methods for human movement dynamics with emphasis on gait. Multibody System Dynamics 47, 265–292. https://doi.org/10.1007/s11044-019-09685-1

Falisse, A., Serrancolí, G., Dembia, C.L., Gillis, J., Jonkers, I., de Groote, F., 2019. Rapid predictive simulations with complex musculoskeletal models suggest that diverse healthy and pathological human gaits can emerge from similar control strategies. Journal of the Royal Society Interface 16, 20190402. https://doi.org/10.1098/rsif.2019.0402

Farris, D.J., Sawicki, G.S., 2012. The mechanics and energetics of human walking and running: a joint level perspective. Journal of the Royal Society Interface 9, 110–118. https://doi.org/10.1098/rsif.2011.0182

Febrer-Nafría, M., Nasr, A., Ezati, M., Brown, P., Font-Llagunes, J.M., McPhee, J., 2022. Predictive multibody dynamic simulation of human neuromusculoskeletal systems: A review. Multibody System Dynamics. https://doi.org/10.1007/s11044-022-09852-x

Fregly, B.J., Besier, T.F., Lloyd, D.G., Delp, S.L., Banks, S.A., Pandy, M.G., D'Lima, D.D., 2012. Grand challenge competition to predict in vivo knee loads. Journal of Orthopaedic Research 30, 503–513. https://doi.org/10.1002/jor.22023

Fregly, B.J., Reinbolt, J.A., Rooney, K.L., Mitchell, K.H., Chmielewski, T.L., 2007. Design of patient-specific gait modifications for knee osteoarthritis rehabilitation. IEEE Transactions on Biomedical Engineering 54, 1687–1695. https://doi.org/10.1109/TBME.2007.891934

Geijtenbeek, T., 2019. SCONE: Open Source Software for Predictive Simulation of Biological Motion. Journal of Open Source Software 4, 1421. https://doi.org/10.21105/joss.01421

Glaessgen, E., Stargel, D., 2012. The digital twin paradigm for future NASA and U.S. Air Force vehicles. In Proceedings of the 53rd AIAA/ASME/ASCE/AHS/ASC Structures, Structural Dynamics and Materials Conference. Reston. https://doi.org/10.2514/6.2012-1818

Haelewaters, D., Hofmann, T.A., Romero-Olivares, A.L., 2021. Ten simple rules for Global North researchers to stop perpetuating helicopter research in the Global South. PLOS Computational Biology 17, e1009277. https://doi.org/10.1371/journal.pcbi.1009277

Halilaj, E., Rajagopal, A., Fiterau, M., Hicks, J.L., Hastie, T.J., Delp, S.L., 2018. Machine learning in human movement biomechanics: Best practices, common pitfalls, and new opportunities. Journal of Biomechanics 81, 1–11. https://doi.org/10.1016/j.jbiomech.2018.09.009

Haralabidis, N., Saxby, D.J., Pizzolato, C., Needham, L., Cazzola, D., Minahan, C., 2020. Fusing accelerometry with videography to monitor the effect of fatigue on punching performance in elite boxers. Sensors 20, 5749. https://doi.org/10.3390/s20205749

Hargraves, C.R., Paris, S.W., 1987. Direct trajectory optimization using nonlinear programming and collocation. Journal of Guidance, Control, and Dynamics 10, 338–342. https://doi.org/10.2514/3.20223

Hatze, H., 1976. The complete optimization of a human motion. Mathematical Biosciences 28, 99–135. https://doi.org/10.1016/0025-5564(76)90098-5

Hicks, J.L., Boswell, M.A., Althoff, T., Crum, A.J., Ku, J.P., Landay, J.A., Moya, P.M.L., Murnane, E.L., Snyder, M.P., King, A.C., Delp, S.L., 2023. Leveraging mobile technology for public health promotion: A multidisciplinary perspective. Annual Review of Public Health 44. https://doi.org/10.1146/annurev-publhealth-060220-041643

Hicks, J.L., Delp, S.L., Schwartz, M.H., 2011. Can biomechanical variables predict improvement in crouch gait? Gait & Posture 34, 197–201. https://doi.org/10.1016/j.gaitpost.2011.04.009

Hicks, J.L., Uchida, T.K., Seth, A., Rajagopal, A., Delp, S.L., 2015. Is my model good enough? Best practices for verification and validation of musculoskeletal models and simulations of movement. Journal of Biomechanical Engineering 137, 020905. https://doi.org/10.1115/1.4029304

Hoy, M.G., Zajac, F.E., Gordon, M.E., 1990. A musculoskeletal model of the human lower extremity: The effect of muscle, tendon, and moment arm on the moment-angle relationship of musculotendon actuators at the hip, knee, and ankle. Journal of Biomechanics 23, 157–169. https://doi.org/10.1016/0021-9290(90)90349-8

Hutchinson, J.R., Anderson, F.C., Blemker, S.S., Delp, S.L., 2005. Analysis of hindlimb muscle moment arms in Tyrannosaurus rex using a three-dimensional musculoskeletal computer model: Implications for stance, gait, and speed. Paleobiology 31, 676–701. https://doi.org/10.1666/04044.1

Hutchinson, J.R., Rankin, J.W., Rubenson, J., Rosenbluth, K.H., Siston, R.A., Delp, S.L., 2015. Musculoskeletal modelling of an ostrich (*Struthio camelus*) pelvic limb: influence of limb orientation on muscular capacity during locomotion. PeerJ 3, e1001. https://doi.org/10.7717/peerj.1001

Jiang, Y., Ye, Y., Gopinath, D., Won, J., Winkler, A.W., Liu, C.K., 2022. Transformer Inertial Poser: Real-time human motion reconstruction from sparse IMUs with simultaneous terrain generation. In the SIGGRAPH Asia 2022 Conference Papers. New York. https://doi.org/10.1145/3550469.3555428

Johnson, W.L., Jindrich, D.L., Roy, R.R., Edgerton, V.R., 2008. A three-dimensional model of the rat hindlimb: Musculoskeletal geometry and muscle moment arms. Journal of Biomechanics 41, 610–619. https://doi.org/10.1016/j.jbiomech.2007.10.004

Kane, T.R., Levinson, D.A., 1985. Dynamics, Theory and Applications. McGraw-Hill, New York.

Kaplan, M.L., H. Heegaard, J., 2001. Predictive algorithms for neuromuscular control of human locomotion. Journal of Biomechanics 34, 1077–1083. https://doi.org/10.1016/S0021-9290(01)00057-4

Karatsidis, A., Bellusci, G., Schepers, H., de Zee, M., Andersen, M., Veltink, P., 2016. Estimation of ground reaction forces and moments during gait using only inertial motion capture. Sensors 17, 75. https://doi.org/10.3390/s17010075

Kidziński, Ł., Mohanty, S.P., Ong, C., Hicks, J.L., Carroll, S.F., Levine, S., Salathé, M., Delp, S.L., 2018. Learning to Run challenge: Synthesizing physiologically accurate motion using deep reinforcement learning. In: Escalera, S., Weimer, M. (Eds.), The NIPS '17 Competition: Building Intelligent Systems. The Springer Series on Challenges in Machine Learning. Springer, Cham, pp. 101-120. https://doi.org/10.1007/978-3-319-94042-7_6

Killen, B.A., Falisse, A., de Groote, F., Jonkers, I., 2020. In silico-enhanced treatment and rehabilitation planning for patients with musculoskeletal disorders: Can musculoskeletal modelling and dynamic simulations really impact current clinical practice? Applied Sciences 10, 7255. https://doi.org/10.3390/app10207255

King, A.C., Campero, M.I., Sheats, J.L., Castro Sweet, C.M., Hauser, M.E., Garcia, D., Chazaro, A., Blanco, G., Banda, J., Ahn, D.K., Fernandez, J., Bickmore, T., 2020. Effects of counseling by peer human advisors vs computers to increase walking in underserved populations. JAMA Internal Medicine 180, 1481–1490. https://doi.org/10.1001/jamainternmed.2020.4143

Kinney, A.L., Besier, T.F., D'Lima, D.D., Fregly, B.J., 2013. Update on Grand Challenge Competition to Predict in Vivo Knee Loads. Journal of Biomechanical Engineering 135, 021012. https://doi.org/10.1115/1.4023255

Knarr, B.A., Reisman, D.S., Binder-Macleod, S.A., Higginson, J.S., 2013. Understanding compensatory strategies for muscle weakness during gait by simulating activation deficits seen post-stroke. Gait & Posture 38, 270–275. https://doi.org/10.1016/j.gaitpost.2012.11.027

Kuo, A.D., Donelan, J.M., 2010. Dynamic principles of gait and their clinical implications. Physical Therapy 90, 157–174. https://doi.org/10.2522/ptj.20090125

Labarrière, F., Thomas, E., Calistri, L., Optasanu, V., Gueugnon, M., Ornetti, P., Laroche, D., 2020. Machine learning approaches for activity recognition and/or activity prediction in locomotion assistive devices—A systematic review. Sensors 20, 6345. https://doi.org/10.3390/s20216345

Lai, A.K.M., Arnold, A.S., Wakeling, J.M., 2017. Why are antagonist muscles co-activated in my simulation? A musculoskeletal model for analysing human locomotor tasks. Annals of Biomedical Engineering 45, 2762–2774. https://doi.org/10.1007/s10439-017-1920-7

Liu, M.Q., Anderson, F.C., Schwartz, M.H., Delp, S.L., 2008. Muscle contributions to support and progression over a range of walking speeds. Journal of Biomechanics 41, 3243–3252. https://doi.org/10.1016/j.jbiomech.2008.07.031

Maas, S.A., Ellis, B.J., Ateshian, G.A., Weiss, J.A., 2012. FEBio: Finite Elements for Biomechanics. Journal of Biomechanical Engineering 134, 011005. https://doi.org/10.1115/1.4005694

Mahmood, N., Ghorbani, N., Troje, N.F., Pons-Moll, G., Black, M.J., 2019. AMASS: Archive of Motion Capture as Surface Shapes. arXiv. https://doi.org/10.48550/arXiv.1904.03278

Mansouri, M., Reinbolt, J.A., 2012. A platform for dynamic simulation and control of movement based on OpenSim and MATLAB. Journal of Biomechanics 45, 1517–1521. https://doi.org/10.1016/j.jbiomech.2012.03.016

McFarland, D.C., McCain, E.M., Poppo, M.N., Saul, K.R., 2019. Spatial dependency of glenohumeral joint stability during dynamic unimanual and bimanual pushing and pulling. Journal of Biomechanical Engineering 141, 051006. https://doi.org/10.1115/1.4043035

Monu, U.D., Jordan, C.D., Samuelson, B.L., Hargreaves, B.A., Gold, G.E., McWalter, E.J., 2017. Cluster analysis of quantitative MRI T$_2$ and T$_{1\rho}$ relaxation times of cartilage identifies differences between healthy and ACL-injured individuals at 3T. Osteoarthritis and Cartilage 25, 513–520. https://doi.org/10.1016/j.joca.2016.09.015

Mortensen, J.D., Vasavada, A.N., Merryweather, A.S., 2018. The inclusion of hyoid muscles improve moment generating capacity and dynamic simulations in musculoskeletal models of the head and neck. PLOS ONE 13, e0199912. https://doi.org/10.1371/journal.pone.0199912

Neptune, R.R., Kautz, S.A., Zajac, F.E., 2001. Contributions of the individual ankle plantar flexors to support, forward progression and swing initiation during walking. Journal of Biomechanics 34, 1387–1398. https://doi.org/10.1016/S0021-9290(01)00105-1

O'Day, J., Lee, M., Seagers, K., Hoffman, S., Jih-Schiff, A., Kidziński, Ł., Delp, S., Bronte-Stewart, H., 2022. Assessing inertial measurement unit locations for freezing of gait detection and patient preference. Journal of NeuroEngineering and Rehabilitation 19, 20. https://doi.org/10.1186/s12984-022-00992-x

O'Neill, M.C., Lee, L.-F., Larson, S.G., Demes, B., Stern, J.T., Umberger, B.R., 2013. A three-dimensional musculoskeletal model of the chimpanzee (*Pan troglodytes*) pelvis and hind limb. Journal of Experimental Biology 216, 3709–3723. https://doi.org/10.1242/jeb.079665

Ong, C.F., Geijtenbeek, T., Hicks, J.L., Delp, S.L., 2019. Predicting gait adaptations due to ankle plantarflexor muscle weakness and contracture using physics-based musculoskeletal simulations. PLOS Computational Biology 15, e1006993. https://doi.org/10.1371/journal.pcbi.1006993

Pellikaan, P., Giarmatzis, G., vander Sloten, J., Verschueren, S., Jonkers, I., 2018. Ranking of osteogenic potential of physical exercises in postmenopausal women based on femoral neck strains. PLOS ONE 13, e0195463. https://doi.org/10.1371/journal.pone.0195463

Piazza, S.J., 2006. Muscle-driven forward dynamic simulations for the study of normal and pathological gait. Journal of NeuroEngineering and Rehabilitation 3, 5. https://doi.org/10.1186/1743-0003-3-5

Pitto, L., Kainz, H., Falisse, A., Wesseling, M., van Rossom, S., Hoang, H., Papageorgiou, E., Hallemans, A., Desloovere, K., Molenaers, G., van Campenhout, A., de Groote, F., Jonkers, I., 2019. SimCP: A simulation platform to predict gait performance following orthopedic intervention in children with cerebral palsy. Frontiers in Neurorobotics 13, 54. https://doi.org/10.3389/fnbot.2019.00054

Pizzolato, C., Lloyd, D.G., Sartori, M., Ceseracciu, E., Besier, T.F., Fregly, B.J., Reggiani, M., 2015. CEINMS: A toolbox to investigate the influence of different neural control solutions on the prediction of muscle excitation and joint moments during dynamic motor tasks. Journal of Biomechanics 48, 3929–3936. https://doi.org/10.1016/j.jbiomech.2015.09.021

Pizzolato, C., Reggiani, M., Saxby, D.J., Ceseracciu, E., Modenese, L., Lloyd, D.G., 2017. Biofeedback for gait retraining based on real-time estimation of tibiofemoral joint contact forces. IEEE Transactions on Neural Systems and Rehabilitation Engineering 25, 1612–1621. https://doi.org/10.1109/TNSRE.2017.2683488

Pizzolato, C., Saxby, D.J., Palipana, D., Diamond, L.E., Barrett, R.S., Teng, Y.D., Lloyd, D.G., 2019. Neuromusculoskeletal modeling-based prostheses for recovery after spinal cord injury. Frontiers in Neurorobotics 13, 97. https://doi.org/10.3389/fnbot.2019.00097

Raissi, M., Perdikaris, P., Karniadakis, G.E., 2019. Physics-informed neural networks: A deep learning framework for solving forward and inverse problems involving nonlinear partial differential equations. Journal of Computational Physics 378, 686–707. https://doi.org/10.1016/j.jcp.2018.10.045

Rajagopal, A., Dembia, C.L., DeMers, M.S., Delp, D.D., Hicks, J.L., Delp, S.L., 2016. Full-body musculoskeletal model for muscle-driven simulation of human gait. IEEE Transactions on Biomedical Engineering 63, 2068–2079. https://doi.org/10.1109/TBME.2016.2586891

Rajagopal, A., Kidziński, Ł., McGlaughlin, A.S., Hicks, J.L., Delp, S.L., Schwartz, M.H., 2020. Pre-operative gastrocnemius lengths in gait predict outcomes following gastrocnemius lengthening surgery in children with cerebral palsy. PLOS ONE 15, e0233706. https://doi.org/10.1371/journal.pone.0233706

Rankin, J.W., Rubenson, J., Hutchinson, J.R., 2016. Inferring muscle functional roles of the ostrich pelvic limb during walking and running using computer optimization. Journal of the Royal Society Interface 13, 20160035. https://doi.org/10.1098/rsif.2016.0035

Ratner, A., Bach, S.H., Ehrenberg, H., Fries, J., Wu, S., Ré, C., 2017. Snorkel. Proceedings of the VLDB Endowment 11, 269–282. https://doi.org/10.14778/3157794.3157797

Rosenberg, M., Steele, K.M., 2017. Simulated impacts of ankle foot orthoses on muscle demand and recruitment in typically-developing children and children with cerebral palsy and crouch gait. PLOS ONE 12, e0180219. https://doi.org/10.1371/journal.pone.0180219

Roupa, I., da Silva, M.R., Marques, F., Gonçalves, S.B., Flores, P., da Silva, M.T., 2022. On the modeling of biomechanical systems for human movement analysis: A narrative review. Archives of Computational Methods in Engineering 29, 4915–4958. https://doi.org/10.1007/s11831-022-09757-0

Sartori, M., Durandau, G., Došen, S., Farina, D., 2018. Robust simultaneous myoelectric control of multiple degrees of freedom in wrist-hand prostheses by real-time neuromusculoskeletal modeling. Journal of Neural Engineering 15, 066026. https://doi.org/10.1088/1741-2552/aae26b

Saul, K.R., Hu, X., Goehler, C.M., Vidt, M.E., Daly, M., Velisar, A., Murray, W.M., 2015. Benchmarking of dynamic simulation predictions in two software platforms using an upper limb musculoskeletal model. Computational Methods in Biomechanics and Biomedical Engineering 18, 1445–1458. https://doi.org/10.1080/10255842.2014.916698

Scherpereel, K.L., Bolus, N.B., Jeong, H.K., Inan, O.T., Young, A.J., 2021. Estimating knee joint load using acoustic emissions during ambulation. Annals of Biomedical Engineering 49, 1000–1011. https://doi.org/10.1007/s10439-020-02641-7

Seireg, A., Arvikar, R.J., 1973. A mathematical model for evaluation of forces in lower extremities of the musculo-skeletal system. Journal of Biomechanics 6, 313–326. https://doi.org/10.1016/0021-9290(73)90053-5

Seth, A., Hicks, J.L., Uchida, T.K., Habib, A., Dembia, C.L., Dunne, J.J., et al., 2018. OpenSim: Simulating musculoskeletal dynamics and neuromuscular control to study human and animal movement. PLOS Computational Biology 14, e1006223.

https://doi.org/10.1371/journal.pcbi.1006223

Sharifi Renani, M., Eustace, A.M., Myers, C.A., Clary, C.W., 2021. The use of synthetic IMU signals in the training of deep learning models significantly improves the accuracy of joint kinematic predictions. Sensors 21, 5876. https://doi.org/10.3390/s21175876

Shimada, S., Golyanik, V., Xu, W., Theobalt, C., 2020. PhysCap. ACM Transactions on Graphics 39, 1–16. https://doi.org/10.1145/3414685.3417877

Shin, C.S., Chaudhari, A.M., Andriacchi, T.P., 2007. The influence of deceleration forces on ACL strain during single-leg landing: A simulation study. Journal of Biomechanics 40, 1145–1152. https://doi.org/10.1016/j.jbiomech.2006.05.004

Shull, P.B., Jirattigalachote, W., Hunt, M.A., Cutkosky, M.R., Delp, S.L., 2014. Quantified self and human movement: A review on the clinical impact of wearable sensing and feedback for gait analysis and intervention. Gait & Posture 40, 11–19. https://doi.org/10.1016/j.gaitpost.2014.03.189

Slade, P., Habib, A., Hicks, J.L., Delp, S.L., 2022. An open-source and wearable system for measuring 3D human motion in real-time. IEEE Transactions on Biomedical Engineering 69, 678–688. https://doi.org/10.1109/TBME.2021.3103201

Song, S., Kidziński, Ł., Peng, X.B., Ong, C., Hicks, J., Levine, S., Atkeson, C.G., Delp, S.L., 2021. Deep reinforcement learning for modeling human locomotion control in neuromechanical simulation. Journal of NeuroEngineering and Rehabilitation 18, 126. https://doi.org/10.1186/s12984-021-00919-y

Stanev, D., Filip, K., Bitzas, D., Zouras, S., Giarmatzis, G., Tsaopoulos, D., Moustakas, K., 2021. Real-time musculoskeletal kinematics and dynamics analysis using marker- and IMU-based solutions in rehabilitation. Sensors 21, 1804. https://doi.org/10.3390/s21051804

Stark, H., Fischer, M.S., Hunt, A., Young, F., Quinn, R., Andrada, E., 2021. A three-dimensional musculoskeletal model of the dog. Scientific Reports 11, 11335. https://doi.org/10.1038/s41598-021-90058-0

Steele, K.M., DeMers, M.S., Schwartz, M.H., Delp, S.L., 2012a. Compressive tibiofemoral force during crouch gait. Gait & Posture 35, 556–560. https://doi.org/10.1016/j.gaitpost.2011.11.023

Steele, K.M., van der Krogt, M.M., Schwartz, M.H., Delp, S.L., 2012b. How much muscle strength is required to walk in a crouch gait? Journal of Biomechanics 45, 2564–2569. https://doi.org/10.1016/j.jbiomech.2012.07.028

Tunca, C., Salur, G., Ersoy, C., 2020. Deep learning for fall risk assessment with inertial sensors: Utilizing domain knowledge in spatio-temporal gait parameters. IEEE Journal of Biomedical and Health Informatics 24, 1994–2005. https://doi.org/10.1109/JBHI.2019.2958879

Uchida, T.K., Seth, A., Pouya, S., Dembia, C.L., Hicks, J.L., Delp, S.L., 2016. Simulating ideal assistive devices to reduce the metabolic cost of running. PLOS ONE 11, e0163417. https://doi.org/10.1371/journal.pone.0163417

Uhlrich, S.D., Falisse, A., Kidziński, Ł., Muccini, J., Ko, M., Chaudhari, A.S., Hicks, J.L., Delp, S.L., 2022a. OpenCap: 3D human movement dynamics from smartphone videos. bioRxiv. https://doi.org/10.1101/2022.07.07.499061

Uhlrich, S.D., Jackson, R.W., Seth, A., Kolesar, J.A., Delp, S.L., 2022b. Muscle coordination retraining inspired by musculoskeletal simulations reduces knee contact force. Scientific Reports 12, 9842. https://doi.org/10.1038/s41598-022-13386-9

Umberger, B.R., 2010. Stance and swing phase costs in human walking. Journal of the Royal Society Interface 7, 1329–1340. https://doi.org/10.1098/rsif.2010.0084

van den Bogert, A.J., Geijtenbeek, T., Even-Zohar, O., Steenbrink, F., Hardin, E.C., 2013. A real-time system for biomechanical analysis of human movement and muscle function.

Medical & Biological Engineering & Computing 51, 1069–1077.
https://doi.org/10.1007/s11517-013-1076-z

van Rossom, S., Smith, C.R., Thelen, D.G., Vanwanseele, B., van Assche, D., Jonkers, I., 2018. Knee joint loading in healthy adults during functional exercises: Implications for rehabilitation guidelines. Journal of Orthopaedic & Sports Physical Therapy 48, 162–173. https://doi.org/10.2519/jospt.2018.7459

van Veen, B., Montefiori, E., Modenese, L., Mazzà, C., Viceconti, M., 2019. Muscle recruitment strategies can reduce joint loading during level walking. Journal of Biomechanics 97, 109368. https://doi.org/10.1016/j.jbiomech.2019.109368

Vanlandewijck, Y., Theisen, D., Daly, D., 2001. Wheelchair propulsion biomechanics. Sports Medicine 31, 339–367. https://doi.org/10.2165/00007256-200131050-00005

Walter, J.P., D'Lima, D.D., Colwell, C.W., Fregly, B.J., 2010. Decreased knee adduction moment does not guarantee decreased medial contact force during gait. Journal of Orthopaedic Research 28, 1348–1354. https://doi.org/10.1002/jor.21142

Werling, K., Raitor, M., Stingel, J., Hicks, J.L., Collins, S., Delp, S.L., Liu, C.K., 2022. Rapid bilevel optimization to concurrently solve musculoskeletal scaling, marker registration, and inverse kinematic problems for human motion reconstruction. bioRxiv. https://doi.org/10.1101/2022.08.22.504896

Wickiewicz, T.L., Roy, R.R., Powell, P.L., Edgerton, V.R., 1983. Muscle architecture of the human lower limb. Clinical Orthopaedics and Related Research 179, 275–283. https://doi.org/10.1097/00003086-198310000-00042

Willson, A.M., Richburg, C.A., Czerniecki, J., Steele, K.M., Aubin, P.M., 2020. Design and development of a quasi-passive transtibial biarticular prosthesis to replicate gastrocnemius function in walking. Journal of Medical Devices 14, 025001. https://doi.org/10.1115/1.4045879

Yamaguchi, G.T., Zajac, F.E., 1990. Restoring unassisted natural gait to paraplegics via functional neuromuscular stimulation: A computer simulation study. IEEE Transactions on Biomedical Engineering 37, 886–902. https://doi.org/10.1109/10.58599

Yamaguchi, G.T., Zajac, F.E., 1989. A planar model of the knee joint to characterize the knee extensor mechanism. Journal of Biomechanics 22, 1–10. https://doi.org/10.1016/0021-9290(89)90179-6

Yuan, Y., Wei, S.-E., Simon, T., Kitani, K., Saragih, J., 2021. SimPoE: Simulated character control for 3D human pose estimation. In Proceedings of the 2021 IEEE/CVF Conference on Computer Vision and Pattern Recognition (CVPR). Nashville. 10.1109/CVPR46437.2021.00708

Zhang, J., Zhao, Y., Shone, F., Li, Z., Frangi, A.F., Xie, S.Q., Zhang, Z.-Q., 2023. Physics-informed deep learning for musculoskeletal modeling: Predicting muscle forces and joint kinematics from surface EMG. IEEE Transactions on Neural Systems and Rehabilitation Engineering 31, 484–493. https://doi.org/10.1109/TNSRE.2022.3226860

# Conflict of interest statement